\documentclass[12pt,notitlepage]{article}
\usepackage[latin9]{inputenc}
\usepackage[a4paper]{geometry}
\usepackage{amsmath}
\usepackage{amssymb}
\usepackage[unicode=true,
 bookmarks=false,
 breaklinks=false,pdfborder={0 0 1},backref=section,colorlinks=false]
 {hyperref}

\makeatletter

\usepackage{units}
\usepackage{units}
\usepackage{xcolor}
\usepackage{graphicx}
\usepackage{ulem}
\usepackage{amsfonts}
\usepackage{caption}
\usepackage[singlespacing]{setspace}
\usepackage{cite}
\providecommand{\U}[1]{\protect\rule{.1in}{.1in}}

\providecolor{lyxadded}{rgb}{0,0,1}
\providecolor{lyxdeleted}{rgb}{1,0,0}

\DeclareRobustCommand{\lyxsout}[1]{\ifx\\#1\else\sout{#1}\fi}

\makeatother

\begin{document}
\title{Constraints on extra dimensions theories from gravitational quantum
barrier experiments}
\author{J. M. Rocha and F. Dahia\\
{\small{}Department of Physics, Federal University of Paraíba - João
Pessoa - PB - Brazil.}}
\maketitle
\begin{abstract}
We discuss the quantum-bouncer experiment involving ultracold neutrons
in a braneworld scenario. Extra-dimensional theories typically predict
the strengthening of gravitational interactions over short distances.
In this paper, we specifically study the anomalous gravitational interaction
between the bouncing neutron and the reflecting mirror, resulting
from hidden dimensions, and its effect on the outcome of this experiment
in the context of a thickbrane model. This analysis allows us to identify
which physical quantity of this extra-dimensional theory this neutron
experiment is capable of constraining.

Based on the experimental data, we found a new and independent empirical
bound on free parameters of the model: the higher-dimensional gravitational
constant and a parameter related to a transverse width of the confined
matter inside the thickbrane. This new bound is valid in scenarios
with an arbitrary number of extra dimensions greater than two. In
this manner, by considering the thickness of the brane, we have been
able to extend previous studies on this topic, which were limited
to models with few codimensions, due to non-computability problems
of power-law corrections of the gravitational potential.
\end{abstract}

\section{Introduction}

Extra dimensions have become, over the years, an important conceptual
ingredient in the search for unified theories in Physics since the
advent of the Kaluza-Klein theory. The existence of hidden dimensions
is a very intriguing theoretical possibility that, more recently,
has also attracted the attention of many experimental research teams,
mainly due to the formulation of the so-called large extra dimension
theories \cite{ADD,add2,rs1,rs2}. As part of the braneworld scenario,
these theories assume that matter and fields are trapped in a sub-space
of three spatial dimensions (3-brane), and only the gravitational
field is capable of propagating in all directions. Because of this
peculiarity, the size of compact hidden dimensions could, from a phenomenological
point of view, be much greater than the Planck length, as taken by
the older Kaluza-Klein theories. Thus, in this framework, the gravitational
interaction may be modified in length scales that would, in principle,
be empirically accessible.

The key ideas of the large extra dimensions theories were introduced
by the so-called ADD model \cite{ADD}, which was originally conceived
as an alternative explanation for the hierarchy problem. According
to this model, the spacetime has a certain number $\delta$ of additional
compact dimensions with a characteristic radius $R$. The dispersion
of the gravitational field along these extra directions would be the
reason for the huge disparity between the strength of gravity and
the other forces of nature when the gravitational field is measured
over distances larger than $R$.

Another interesting feature of this model is that, on the other side,
it predicts that the gravitational force between two massive particles
at a distance $r$ much shorter than $R$ will be amplified by a factor
proportional to $(R/r)$$^{\delta}$, when compared to the Newtonian
gravitational force. This hypothetical magnification of the gravitational
interaction, which is also predicted by other versions of extra-dimensions
theories \cite{add2,rs1,rs2}, has stimulated the search for deviations
of the inverse square law in ever smaller domains. Since then, many
theoretical and experimental works have been conducted with the intention
of detecting signs of the existence of such extra dimensions in diverse
branches of physics \cite{Murata,Adelberger2003}, ranging from cosmology
and astrophysics \cite{pdgAlex} to high-energy particle collisions
\cite{pdgAlex,colliders}. There are also many laboratory tests pursuing
this same goal. They are based on a wide variety of experiments involving
torsion balance \cite{Adelberger}, the Casimir effect \cite{Casimir,Casimir1,Casimir2,Casimir3},
spectroscopy \cite{atomicspec1,atomicspec2,atomicspec3,molecule,Fabio2016,safronova,Fabio2019,Adiel}
and neutron interferometry \cite{Greene,Nesvizhevsky2008,Jesriel},
just to mention a few examples.

Often, in these approaches, the Yukawa parameterization for the potential
energy is used, where deviations are described by the factor $\alpha e^{-r/\lambda}$,
containing two free parameters $\alpha$ and $\lambda$, which refer
to the intensity and the range of the anomalous interaction, respectively.
Experimental data from different systems impose limits on the parameters
$\alpha$ and $\lambda$ on diverse length scale ranges \cite{Murata,Casimir3,revision}.

The Yukawa parameterization is very useful since it allows us to consider
modifications of gravitation that may have different motivations and
theoretical origins \cite{extensions,models}, as for example, in
some models of the F(R) type \cite{stelle}. In the braneworld scenario,
this Yukawa potential would describe the correction of the Newtonian
potential for a setting in which the interacting particles are far
apart compared to the compactification radius. In this case, the parameter
$\alpha$ would give us some information about the number of extra
dimensions and $\lambda$ would be of the order of $R$ \cite{kehagias}.

However, when the interacting particles are sufficiently close or
when their wave functions overlap, the appropriate correction for
the gravitational potential is the power-law type, as previously mentioned.
This occurs in some physical systems, such as in the neutron interferometry,
when the particle passes trough a material body, which is acting as
a phase shifter \cite{Jesriel}. Under these circumstances, we have
to deal with the fact the gravitational potential inside a material
is not computable when there are two or more additional dimensions.
In these cases, it is necessary to consider the brane thickness \cite{Rubakov}
to avoid divergence problems and thus be able to investigate the effects
of the hidden dimensions on that system.

In this context, neutron experiments offer special conditions for
testing gravity below the sub-millimeter scale \cite{Abele,Snow},
since the interference of electromagnetic interactions is relatively
low. This property allows neutrons to penetrate deeper into matter,
revealing how this particle interacts with the atomic nucleus. The
analysis of neutron-nuclei collisions could, in principle, provide
some information about speculative departures of the usual gravitational
interaction at nucleus length scale \cite{Abele,Frank}. This has
been investigated by exploring several neutron scattering experiments
\cite{Frank,leeb,kamiya,Haddock}.

In our present study here, we shall consider another neutron experiment
that was designed to study the behavior of neutrons under the influence
of the Earth's gravitational field acting as a potential barrier for
the particle's vertical movement \cite{Nesvizhevsky2002,Nesvizhevsky2005,Westphal}.
This experiment has been utilized extensively in the search for signals
of modified gravity at the microscopic scale \cite{Abele,Abele2003,Nesvizhevsky2004,Westphal2,Fabien}.
Basically, it consists of a beam of ultra-cold neutrons that moves
freely in the horizontal direction inside an apparatus, while its
vertical motion is bound by the Earth's gravitational field and by
a reflecting surface in the bottom. From a theoretical point of view,
the neutron could be idealized, regarding its vertical motion, as
a quantum bouncer \cite{Nesvizhevsky2002}. This system is known to
have a discrete set of bound states, each one associated with a characteristic
height. The possibility of finding neutrons in these quantum states
induced by the Earth's gravitational field was experimentally examined
and discussed in Refs.\cite{Nesvizhevsky2002,Westphal}. Placing an
absorber above the mirror, the neutron transmission rate was measured
by a detector located at the end of the device. The transmission rate
is a function of the slit size between the absorber and the mirror,
which could be varied. According to Ref. \cite{Westphal}, the data
are consistent with the quantum bouncer model.

As the corresponding height of the lowest energy levels is of a few
micrometers, this experiment has been widely employed to examine diverse
aspects of gravity in this length scale \cite{koch}, including even
discussions about an entropic origin of gravity \cite{entropic1,entropic2}.
Another interesting application, as we have already mentioned, is
to investigate any change of inverse square law of the Newtonian gravitational
force in this domain by considering the interaction between the neutron
and the mirror \cite{Abele,Abele2003,Nesvizhevsky2004,Westphal2,Fabien}.

The reflection of the neutron by the mirror is due to the collective
interaction between the neutron and the nuclei of the atoms in the
mirrror via the nuclear force. Under the conditions of the experiment,
this interaction can be described by the nuclear Fermi pseudo-potential.
Of course, the neutron will also interact with the glass plate (the
mirror) through the gravitational interaction too. The classical Newtonian
interaction is negligible, but a speculative, short range amplification
of the gravitational interaction can be explored on the basis of this
experiment \cite{Abele,Abele2003,Nesvizhevsky2004,Westphal2,Fabien}.

Previous analyses of this experiment found that the consistency between
the Standard Model predictions and the empirical data demands that
$\alpha<1\times10^{15}$ for $\lambda=1$$\mu m$, for example, when
the deviations are described by the Yukawa parameterization \cite{Nesvizhevsky2004,Westphal2}.
The data from this experiment have also been analyzed taking into
account the power-law correction for the neutron-mirror gravitational
interaction \cite{Fabien}. Empirical limits for the power-law strengthening
of the gravity were found only for $\delta\leq4$ \cite{Fabien}.
Scenarios with more codimensions could not be appropriately investigated
due to the divergence problems in the calculation of the potential
energy of the neutron-mirror gravitational interaction in the higher-dimensional
scenario. At a first sight, as pointed out in Ref. \cite{Fabien},
power-law corrections with $\delta>4$ seem to be incompatible with
neutron bound states.

In this paper, however, we show that it is possible to extend the
analysis of this experiment to scenarios with any number of extra-dimensions.
As we shall see, in a thickbrane model, we can explicitly calculate
the gravitational potential produced by the mirror in the leading
order, overcoming the divergence problems. Consequently, we can identify
the quantity that can be constrained by the data from this neutron
experiment, even when the space has five or more additional extra-dimensions.
As we shall see this quantity depends on the gravitational constant
of the higher-dimensional space and on a parameter related to the
confinement width of the matter inside the thickbrane.

Although the constrains obtained in this paper are not as strong as
empirical bounds for a similar quantity extracted from colliders data,
for example, our work fills a gap in the literature, since there were
no experimental limits to the power-law correction for the gravitational
interaction in the case in which the number of extra dimensions is
greater than four, based on this important neutron experiment. Furthermore,
this work, by demonstrating that the anomalous gravitational potential
energy between the mirror and neutron is finite for any number of
additional dimensions when the brane's thickness is taken into account,
shows that the evidence of bound states for ultracold neutrons does
not rule out extra dimensions theories with five or more hidden dimension
in thickbrane scenarios.

\section{The Quantum Bouncer subjected to an anomalous gravitational interaction}

The quantum-bouncer experiment was designed to investigate whether
the gravitational field of the Earth can induce a quantum bound state
for ultra-cold neutrons \cite{Nesvizhevsky2002}. A beam of neutrons
transverses a slit formed by a mirror at the bottom and an absorber
positioned few micrometers above. The mirror has a length of about
10 cm and the slit size is adjustable \cite{Nesvizhevsky2005}. The
neutrons that manage to enter the slit after passing through a set
of collimators arrive there at speeds of about 5 m/s in the horizontal
direction. The vertical velocity can be controlled by the height of
the absorber. At 125 $\mu$m, for instance, the absorber permits that
only neutrons with vertical speed less than 5 cm/s transverse the
slit without being captured. These neutrons would describe, according
to the classical mechanics, a sequence of parabolic trajectories between
two consecutive collisions with the mirror, due to the action of the
Earth's gravitational force. Along the mirror length, each ultracold
neutron would be reflected by the mirror a dozen times before reaching
the detector at the end of the apparatus.

The mirror is an optical glass block with dimensions 10 cm $\times$10
cm $\times$3 cm \cite{Westphal}. The reflection of the neutron by
the mirror is caused by the nuclear interaction between the neutron
and the silicon dioxide nuclei of the glass, according to the Standard
Model. For ultra-cold neutrons, the interaction with each nuclei can
be described by the Fermi pseudo-potential. Further, as the incident
neutrons has wavelengths much greater the atomic separation in the
glass, the interaction with the mirror can be described by neutron
optical potential, given by: 
\begin{equation}
U_{op}=\frac{2\pi\hslash^{2}}{m}Nb,\label{p.optical}
\end{equation}
where $m$ is the neutron mass, $b$ is the average neutron-nucleus
scattering length and $N$ is the atomic number density of the material.
The scattering length is a phenomenological quantity whose value depends
predominately on the nuclear interaction between the nucleus and the
incident neutron according to the standard model. For the material
used in the experiment, $U_{op}=100$ neV approximately \cite{Nesvizhevsky2005}.

An important characteristic of this experiment is that neutrons hit
the mirror with a very low vertical velocity. The corresponding kinetic
energy is of the order of a few peV. Thus, for those incidents neutrons,
$U_{op}$ works as an impenetrable potential barrier. Therefore, the
potential energy of this system can be written as: 
\begin{equation}
U\left(z\right)=\left\{ \begin{array}{c}
\infty,\text{para }z<0;\\
mgz,\text{ para }z\geq0,
\end{array}\right.
\end{equation}
where $z$ is the height of the neutron with respect to the mirror
and $g$ is the Earth's gravitational acceleration evaluated in the
lab, which is almost constant inside the apparatus.

The motion of the neutron in the vertical direction is, then, governed
by the Schrödinger equation, which, for positive $z$, has the form:
\begin{equation}
-\frac{\hbar^{2}}{2m}\frac{d^{2}\psi}{dz^{2}}+mgz\psi=E\psi,\label{Schrodinger}
\end{equation}
where $E$ is the energy of the neutron minus the kinetic energy associated
to the decoupled horizontal motion.

Usually the equation (\ref{Schrodinger}) is analyzed in terms of
a dimensionless parameter defined as $\zeta=z/z_{0}$, where $z_{0}=\left(\hbar^{2}/2m^{2}g\right)^{1/3}$
corresponds to a natural length scale set by the system. This parameter,
whose value is around 5,8 $\mu m$, may be interpreted as the height
the neutron could reach if its linear momentum had the magnitude $p_{z}\sim\hbar/z_{0}$,
roughly estimated from the Heisenberg uncertainty principle.

The solutions of equation (\ref{Schrodinger}) that satisfies the
appropriate boundary condition can be expressed in terms of the Airy
function as: 
\begin{equation}
\psi_{n}=N_{n}Ai\left(\zeta-\zeta_{n}\right).\label{autofun}
\end{equation}
This solution is valid for $z\geq0$. In the negative side of the
$z-$axis, the wave-function is null. For each allowed $n$ ($=1,2,3,..$),
the corresponding wave-function represents a bound state for the quantum
bouncer. The normalization constants can be expressed in terms of
derivative of the Airy function, $Ai^{\prime}(z)$, in the following
form \cite{Fabien}: 
\begin{equation}
N_{n}=\frac{z_{0}^{-1/2}}{\left\vert Ai^{\prime}\left(-\zeta_{n}\right)\right\vert },
\end{equation}
where, in order to satisfy the boundary condition at $z=0$, the constants
$-\zeta_{n}$ should be roots of the Airy function. This last condition
implies that the quantum bouncer's energy eigenvalues constitute a
discrete set, which can be written as $E_{n}=mgz_{0}\zeta_{n}$. If
the factor $z_{n}=z_{0}\zeta_{n}$ is interpreted as classical return
points, a pictorial view of this discretization is that, in the quantum
regime, the bouncing particles are only allowed to jump up to permissible
heights $z_{n}$.

This behavior, of course, has consequences on the neutron transmission
rate through the slit formed by the mirror and the absorber. For example,
we may expect that, when the slit is narrower than $z_{1}$ (the lowest
return point $z_{0}\zeta_{1}=13,7$$\mu m$) no neutron could reach
the detector. This contradicts the classical prediction. Indeed, if
the incident flux is spatially uniform and neutron's vertical velocity
is also distributed uniformly in the beam, then, according to classical
mechanics, it is expected that the detected neutrons are only those
which are found inside the slit (of size $h,$ let us say) with vertical
velocity less than $v_{max}=(2gh)^{1/2}$. Therefore, the count rate
of neutrons at detector should be proportional to the factor $h^{3/2}$
for any $h$.

These different predictions were experimentally examined and the data
turned out to be in favor of quantum bouncer model \cite{Nesvizhevsky2005,Westphal}.
According to Ref. \cite{Nesvizhevsky2005}, for the first state, $n=1$,
the corresponding height $h_{1}$ is given by 
\begin{equation}
h_{1}^{\exp}=12,2\mu m\pm\left(1,8_{syst}+0,7_{stat}\right)\mu m,\label{altura1-1}
\end{equation}
which is in good agreement with the theoretical height for $h_{1}$
which is $h_{1}^{teo}=13,7\mu m$ \cite{Nesvizhevsky2005}. This coincidence
at 1$\sigma$ statistic level can be used to put empirical limits
on an anomalous gravitational interaction that hypothetically acts
in the system. Clearly, the ordinary gravitational interaction between
the neutron and the mirror can be safely ignored in this experiment.
However, an anomalous interaction could, in principle, be much stronger
in the micrometer length scale. Admitting this possibility, the energy
potential would be modified by a new term: 
\begin{equation}
U_{G}=mgz+U_{A}(z).
\end{equation}

This extra energy potential $U_{A}$, associated to the modified gravitational
interaction between the neutron and mirror, introduces corrections
on the energy eigenvalues of the quantum bouncer, and, consequently,
on the theoretical height the bound neutrons could reach. Of course,
due to the compatibility between the experimental value and the theoretical
prediction based on the standard physics framework, the correction
resulting from the additional interaction cannot be greater than the
experimental error.

In next section we are going to determine the new energy potential
$U_{A}$, considering that the origin of the anomalous interaction
is the existence of hidden dimensions in the braneworld scenario.

\section{Modified Gravitational Potential in Thick Branes}

According to the ADD model \cite{ADD}, the space-time has a certain
number of extra dimensions arranged in a topology of a torus. In the
absence of matter, the metric in the bulk and the induced metric in
the brane are assumed to be of Minkowski type. In addition, it is
admitted that the compactification radius has an uniform value $R$
over the entire manifold in this background configuration.

The presence of matter in the brane changes the spacetime geometry,
and the gravitational effects it produces depends on the energy-momentum
distribution. For confined fields, the energy and momentum are concentrated
around the brane, and, therefore, it is reasonable to assume that
its spatial distribution can be described by a tensor with the following
form \cite{colliders}: 
\begin{equation}
T_{AB}=\eta_{A}^{\mu}\eta_{B}^{\nu}T_{\mu\nu}\left(x\right)f\left(w\right).
\end{equation}
Here Greek indices range from $0$ to $3$ and are exclusively associated
to brane's parallel directions, while capital Latin indices range
from $0$ to $3+\delta$ and, therefore, can refer to any direction
of the whole space. The points in the 3-brane are localized by $x-$coordinates
and $w$ gives the position in the extra-space. The spacetime Minkowski
metric is given by $\eta_{AB}$ and $T_{\mu\nu}\left(x\right)$ is
the conventional energy-momentum tensor defined in the brane. The
function $f\left(w\right)$ delineates the energy distribution in
the transverse direction of the brane. If we consider a brane with
zero thickness, $f\left(w\right)$ is a Dirac delta function, however,
for a brane with a non-zero thickness, $f\left(w\right)$ is some
regular and normalized function.

In the weak-field regime, the metric can be written as $g_{AB}=\eta_{AB}+h_{AB}$,
where the perturbation $h_{AB}$ obeys Einstein's linearized equations:
\begin{equation}
\square h_{AB}=-\frac{16\pi G_{D}}{c^{4}}\bar{T}_{AB},\label{Ein.linear}
\end{equation}
in the so-called harmonic gauge. The box is the D'lambertian operator
associated to the Minkowski metric, whose signature is assumed to
be $(-,+...,+)$. The source term is given by: 
\begin{equation}
\bar{T}_{AB}=T_{AB}-\frac{1}{\delta+2}\eta_{AB}T_{C}^{C},
\end{equation}
and the extra-dimensional gravitational constant $G_{D}$ is equal
to the Newtonian gravitational constant multiplied by the extra-space
background volume, i.e., $G_{D}=G\left(2\pi R\right)^{\delta}$ \cite{colliders}.

The exact solution of the equation (\ref{Ein.linear}) is known and
it depends on the topology of the supplementary space \cite{Fabio2016}.
For short distances, however, we can take only the dominant term,
which, in the static regime, is given by 
\begin{equation}
h_{AB}\left(\vec{X}\right)=\frac{16\pi\Gamma\left(\frac{\delta+3}{2}\right)G_{D}}{\left(\delta+1\right)2\pi^{\left(\delta+3\right)/2}c^{4}}\int\frac{\bar{T}_{AB}\left(\vec{X}^{\prime}\right)}{\left\vert \overrightarrow{X}-\vec{X}\right\vert ^{1+\delta}}d^{3+\delta}X^{\prime},\label{h.sol}
\end{equation}
where $\vec{X}$ and $\vec{X}^{\prime}$ are coordinates of the ambient
space. As discussed in \cite{Jesriel}, at this order of approximation,
the potential is not sensible to details of the topology of the extra-dimensional
space. It is well known that the effect of the torus topology on the
solution can be simulated by topological mirror images of the source
spread over the covering space of the torus \cite{Fabio2016,kehagias}.
As the solution (\ref{h.sol}) ignores the potential produced by these
topological images, which would add to the potential (\ref{h.sol}),
then, the metric correction given by (\ref{h.sol}) is a conservative
approximation. As a consequence, the bounds we obtain in this paper
are less stringent than those that we would find from the full solution.
However this approximation is sufficient for our purpose of finding
a finite empirical limit for the free parameters of the model at the
leading order for any number of extra dimensions, complementing, thus,
previous studies whose analyses are valid only for $\delta<5$ \cite{Fabien}.

With the solution (\ref{h.sol}), we can calculate the gravitational
potential produced by the mirror, since in the weak field regime,
valid for the present discussion, we have $\varphi=-c^{2}h_{00}/2.$
The potential energy of the gravitational interaction between a non-relativistic
neutron and the mirror is given by $U_{A}=m\varphi$. In the next
section, we are going to calculate the potential produced by the mirror
in the exterior and interior regions separately.

\subsection{The anomalous gravitational potential produced by the mirror}

In this experiment, the mirror can be described, for the purpose of
calculating the potential, as a semi-infinite plate, since its dimensions
are much larger than $z_{0}$ (the length scale pertinent of the quantum
bouncer). In this context, this material is a non-relativistic gravitational
source and, therefore, $T_{00}$ is the only relevant component of
energy-momentum tensor of the medium. If $\varrho$ denotes the mass
density of the mirror then we have, approximately, $T_{00}=c^{2}\varrho(x)f_{N}(w)$,
where $f_{N}(w)$ is the profile of the nucleus's energy distribution
in transverse directions. Taking this into consideration, it follows
from (\ref{h.sol}), after integrating with respect to the horizontal
directions inside the plate, that the neutron-mirror potential energy
at height $z$ above the mirror is :
\begin{equation}
U_{A,ext}\left(z\right)=-\frac{2\pi m\hat{G}_{D}\rho}{(\delta-1)}\int\frac{f_{N}(w^{\prime})dz^{\prime}d^{\delta}w^{\prime}}{\left(\left(z-z^{\prime}\right)^{2}+w^{\prime2}\right)^{\frac{\delta-1}{2}}},\label{p.ext}
\end{equation}
where, for the sake of simplicity, we have defined: 
\begin{equation}
\hat{G}_{D}=\frac{4\Gamma\left(\frac{\delta+3}{2}\right)}{\left(\delta+2\right)\pi^{\left(\delta+1\right)/2}}G_{D}.
\end{equation}
In the integration (\ref{p.ext}), $w^{\prime}-$coordinates span
the supplementary space and the vertical coordinate $z^{\prime}$
covers the block's interior ( $-\infty<z^{\prime}<0$). For confined
matter in the brane, the effective range of $w^{\prime}$ is of the
order of the brane thickness (which is of the order of the matter
confinement parameter $\sigma$ that will be defined more precisely
later in the Eq. (\ref{sigma}) ). Thus, for $z>>\sigma$, the potential
can be directly computed and it does not depend on the function $f\left(w\right)$,
at the leading order. Therefore, in this region, it coincides with
the potential obtained from the zero-width brane description, as expected.
For $\delta>2$, we can write the gravitational potential as \cite{Fabien}
\begin{equation}
\varphi_{ext}\left(z\right)=-\frac{2\pi m\hat{G}_{D}\rho}{\left(\delta-1\right)\left(\delta-2\right)z^{\delta-2}},\text{ for }\delta>2.
\end{equation}

Although this thin-brane potential diverges at $z=0$, the average
value of the associated potential energy is finite, for neutrons in
the quantum-bouncer states given by Eq. (\ref{autofun}), when $\delta<5$.
This happens because, in the vicinity of the surface of the mirror
at $z=0$, the Airy function function expands as: $Ai^{2}\left(z/z_{0}-\zeta_{n}\right)=Ai^{\prime2}\left(-\zeta_{n}\right)\left(z/z_{0}\right)^{2}+O\left(z^{4}\right)$.

However, for the case in which the supplementary space has five or
more extra dimensions, the energy correction due to the anomalous
gravitational interaction cannot be obtained without taking into account
the brane's thickness. In this context, a practical way to proceed
\cite{Jesriel} is just to modify the potential in the vicinity of
the surface by introducing a term $\gamma\sigma$(of the order of
the brane thickness) in the following way: 
\begin{equation}
\varphi_{ext}\left(z\right)=-\frac{2\pi\hat{G}_{D}\rho}{\left(\delta-1\right)\left(\delta-2\right)\left(z+\gamma\sigma\right)^{\delta-2}},\text{ for }\delta>2.
\end{equation}
Then, the dimensionless parameter $\gamma$ is determined from the
continuity condition between the exterior and the interior potentials
at the surface of the mirror.

\subsubsection{Internal Potential}

In the scenario of an infinitely thin brane, the distribution $f\left(w\right)$,
which describes the matter confinement, would be a Dirac delta function.
In this case, as we have already mentioned, the potential (\ref{p.ext})
diverges in any interior point of the mirror (for $\delta>1$), and,
therefore, the interior gravitational energy potential would not be
computable.

In a less idealized framework, where the brane has a thickness, the
baryonic mass of the confined matter is distributed along the extra
dimension according to some non-singluar function $f_{N}\left(w\right)$,
and the produced potential is finite everywhere. At a point $\vec{x}$
in the brane's center ($w=0$), the potential can be evaluated from
the expression: 
\begin{equation}
\varphi_{int}\left(\vec{x}\right)=-\hat{G}_{D}\int\frac{\rho f_{N}\left(w^{\prime}\right)}{\left(\left\vert \vec{x}-\vec{x}^{\prime}\right\vert ^{2}+w^{\prime2}\right)^{\frac{\delta+1}{2}}}d^{3}x^{\prime}d^{\delta}w^{\prime}.
\end{equation}

Considering that the mass density of the material ($\rho$) is almost
constant inside the plate, then, for every point $\vec{x}$ inside
the medium, the integration with respect $\vec{x}^{\prime}$ (in a
3-ball region centered at $\vec{x}$ and with a radius of order of
the length scale of the brane thickness) gives us, in the leading
order \cite{Jesriel}, the following approximation: 
\begin{equation}
\varphi_{int}\left(x\right)\simeq-\beta\frac{G_{D}\rho}{\sigma^{\delta-2}},\label{p.int}
\end{equation}
where $\beta$ is a constant that depends on the number of dimensions:
\begin{equation}
\beta=\frac{4\left(\delta+1\right)}{\left(\delta^{2}-4\right)\pi^{\left(\delta-1\right)/2}},
\end{equation}
and the parameter $\sigma$, related to the transverse distribution
of the confined matter, is defined by: 
\begin{equation}
\frac{1}{\sigma^{\delta-2}}=\Gamma\left(\delta/2\right)\int\frac{f_{N}\left(z\right)}{z^{\delta-2}}d^{\delta}z\label{sigma}
\end{equation}
This result shows that, in this order of approximation, the most relevant
quantity regarding the mass distribution of the nucleus along the
extra directions comes down to the expected value of $z^{2-\delta}$
with respect the distribution $f_{N}(z)$. This quantity is finite
for a regular function $f_{N}\left(z\right)$ and establishes a notion
of an effective transverse distance of the matter distribution to
brane's center. The numerical factor before the integral in (\ref{sigma})
ensures that $\sigma$ is equal to standard deviation of $f_{N}\left(z\right)$,
multiplied by $\sqrt{2}$, for any number of extra dimensions, when
the mass density obeys a normal distribution in the transverse directions.

By comparing the exterior and interior potential solutions, it follows
from the continuity condition at $z=0$, that, in the leading order,
the parameter $\gamma$ is given by 
\begin{equation}
\gamma=\left[\frac{\Gamma\left(\frac{\delta+1}{2}\right)}{\left(\delta-1\right)\pi^{1/2}}\right]^{\frac{1}{\delta-2}},\text{ for }\delta>2.
\end{equation}
From the internal potential (\ref{p.int}), the influence of the new
interaction on the neutron optical potential can be investigated.
For our discussion here, it is important to ensure that the idealization
we have made about the Fermi-potential is still valid even in the
presence of the additional interaction.

The optical potential can be evaluated from the effective scattering
length $b$ of the neutron interacting with the atoms of the material.
Basically, for our purposes here, we may consider: $b=\bar{b}_{N}+\bar{b}_{A}$,
where $\bar{b}_{N}$ is the average of the scattering length associated
with the nuclear interaction between the neutron and the nuclei of
the medium and $b_{A}$ is the scattering length due to the anomalous
gravitational interaction between those particles.

In the Born approximation, the scattering length $b_{A}$, due to
the neutron anomalous interaction with a single atom, can be calculated
from the expression \cite{leeb}: 
\begin{equation}
b_{A}=\frac{m}{2\pi\hbar^{2}}\int m\varphi_{a}\left(\vec{x}\right)d^{3}\vec{x},\label{bA}
\end{equation}
where $\varphi_{a}$ is the anomalous gravitational potential produced
by an atom. Once again, for $\delta>2$, this potential is not computable
in the infinitely thin brane scenario. However, in the thickbrane
scenario, the potential is finite and, in the leading order, has the
same form of (\ref{p.int}), replacing $\rho$ by the nuclear density
\cite{Jesriel}. Thus, in the calculation of (\ref{bA}), the dominant
contribution comes from the nuclear region. As a consequence, $b_{A}$
is proportional to the mass of the atomic nucleus \cite{Jesriel}.
Taking the average with respect to different atoms of the material
and substituting $\bar{b}_{A}$ in the expression (\ref{p.optical}),
we find that contribution of the anomalous interaction to the neutron
optical potential is equal to $U_{A,int}=m\varphi_{int}$, i.e., the
average of the anomalous gravitational potential energy of the neutron
inside the material. Therefore, taking into account (\ref{p.int}),
the neutron optical potential, in this scenario, is given by 
\begin{equation}
U_{op}=\frac{2\pi\hslash^{2}}{m}Nb_{N}-\frac{\beta G_{D}\rho}{\sigma^{\delta-2}}.
\end{equation}

Contributions from other standard neutron interactions, such as the
electron-neutron interactions, are much smaller than the strong nuclear
force \cite{Snow,leeb} and can be neglected here for the sake of
simplicity. In the next section, we are going to discuss the impact
of this hypothetical gravitational potential on the outcome of the
quantum-bouncer experiment.

\section{New Constraints}

It is expected that the anomalous gravitational interaction between
the neutron and the mirror will change the energy of the quantum bouncer
states. Treating this speculative interaction as a small one when
compared to the neutron-earth gravitational interaction, we can estimate
the corrections on the energy levels of the system due to the new
interaction. According to the usual perturbative method, the shift
in the energy is equal to the average value of the potential energy
of the anomalous interaction evaluated in the original eigenstates
(\ref{autofun}). So, considering the potential found in the previous
section, it follows that, to the first order with respect to $G_{D}$,
the energy of the state $n$ will be shifted by the following amount:
\begin{equation}
\langle U_{A}\rangle_{n}=-\frac{2\pi m\hat{G}_{D}\rho}{\left(\delta-1\right)\left(\delta-2\right)}\frac{1}{\left\vert Ai^{\prime}\left(-\zeta_{n}\right)\right\vert ^{2}}\frac{1}{z_{0}^{\delta-2}}\int_{\gamma\sigma/z_{0}}^{\infty}\frac{Ai^{2}\left(\eta-\zeta_{n}-\gamma\sigma/z_{0}\right)}{\left(\eta\right)^{\delta-2}}d\eta.\label{Ua_mean}
\end{equation}
As expected, this is a negative quantity, due to the attractive nature
of gravity. This means that the neutron-mirror gravitational interaction
will be responsible for decreasing the height of the return point
($h_{n}$) of the bouncing neutrons.

To avoid any contradiction with the empirical data, the predicted
reduction cannot exceed the experimental error. Thus, for the first
state ($n=1$), the anomalous interaction cannot produce a variation
of $h_{1}$ greater than 2.5 $\mu m$ (see Eq. \ref{altura1-1}).
This condition requires that $\left\vert \langle U_{A}\rangle_{1}\right\vert <0.3$
peV, establishing thus experimental bounds for the free parameter
of the model. To find it explicitly, it is necessary to evaluate $\langle U_{A}\rangle_{1}$
in terms of $G_{D}$ and $\sigma$, from eq. (\ref{Ua_mean}). Although
the exact value is not known, a rough estimate for an upper limit
for $\langle U_{A}\rangle_{1}$ can be obtained simply by replacing
the Airy function by the unity, as $\left\vert Ai\right\vert ^{2}<1$.
Proceeding in this way, we find for $\delta>3$: 
\begin{equation}
\left\vert \langle U_{A}\rangle_{n}\right\vert <\frac{2\pi\gamma m\hat{G}_{D}\rho}{\left(\delta-1\right)\left(\delta-2\right)\left(\delta-3\right)\left\vert Ai^{\prime}\left(-\zeta_{n}\right)\right\vert ^{2}\left(\gamma\sigma\right)^{\delta-2}}\left(\frac{\sigma}{z_{0}}\right).
\end{equation}
Comparing this result with the internal energy $m\varphi_{int}$,
calculated in (\ref{p.int}), we can see that this upper estimate
is smaller than the internal potential energy by a factor proportional
to $\left(\sigma/z_{0}\right)$, whose order, for realistic branes,
is less than $10^{-12}$. Actually, for $\delta>5,$ we can obtain
a more precise estimate by expanding the wave-function (\ref{autofun})
around the plate surface. We can show that the contribution is weaker
than the internal potential energy by a factor of the order $\left(\sigma/z_{0}\right)^{3}$.

These results suggests that the best empirical constraints this experiment
could provide should be extracted from the analysis of the internal
gravitational potential on the optical potential.

Although $U_{op}$ is still considered a semi-phenomenological quantity,
ab initio calculations of the neutron optical potential obtained recently
have found satisfactory results \cite{abinitio}. This successful
calculation of the neutron optical potential from the first principles
has strengthened the confidence in the standard model of nuclear physics
in this low energy regime too \cite{abinitio}. Therefore, this reinforces
expectations that the nuclear interaction between the nucleus and
neutron is the dominant one, as predicted by the Standard Model, and
therefore $U_{op}$ should be almost equal to the nuclear Fermi potential
$U_{F}$. Based on these considerations, it follows that, with respect
to the optical potential of this system, it is not acceptable for
the hypothetical higher-dimensional gravitational interaction to have
a contribution of the same order of the strong interaction, otherwise
the neutrons would not be reflected by the mirror and, therefore,
there would be no standing wave states in the vertical direction as
verified by the experiment.

Therefore the observation of bound vertical motions of neutrons as
reported in \cite{Westphal} demands an additional constraint for
the anomalous interaction: 
\begin{equation}
\frac{\beta G_{D}m\rho}{\sigma^{\delta-2}}<U_{F}=10^{-7}eV.\label{constraint}
\end{equation}
This constraint imposes new empirical bound for $G_{D}$ and $\sigma$,
the free parameters of the large extra dimension model. Together,
these two parameters define the energy scale $\Lambda=hc\left(\sigma^{\delta-2}/\ell_{D}^{\delta+2}\right)^{1/4},$
where $\ell_{D}$ is the new length scale associated to the higher-dimensional
gravitational constant $G_{D}$, according to the relation $\ell_{D}^{\delta+2}=G_{D}\hbar/c^{3}$.
To compare this constraint with other experimental limits extracted
from other experiments, it is useful to rewrite the anomalous potential
in terms of an effective energy scale defined by those parameters.
Thus, from (\ref{constraint}), we have 
\begin{equation}
\Lambda^{4}>\frac{8\pi^{3/2}\left(\delta+1\right)}{\left(\delta^{2}-4\right)\pi^{\delta/2}}\frac{hc^{7}m\rho}{U_{F}}.\label{lambda}
\end{equation}

This condition is a new and independent constrain imposed by this
neutron experiment, which is valid for $\delta>2$. Taking the glass
density as $\rho\simeq2.5\,g/cm^{3}$ (a reference value) and the
mass of the neutron from 2018 CODATA recommended values, we find the
following limits: $\Lambda>0.94$ GeV ($\delta=3$); $\Lambda>0.69$
GeV ($\delta=4$); $\Lambda>0.54$ GeV ($\delta=5$) and $\Lambda>0.44$
GeV ($\delta=6$).

These are weak constraints compared to those extracted from high-energy
particle collisions. Actually the exact colliders bounds depend on
the center-of-mass energy and the luminosity of the particle beams,
but, in general terms, recent data impose that a similar quantity
(a cut-off parameter) should be larger than Tev order \cite{pankov,lhc}.
Of course, the collider bounds are much stronger in comparison to
the one set by the inequality (\ref{lambda}). However we have to
emphasize that, in the quantum bouncer experiment, the energy of the
incident neutrons is of the order of peV, so the model of large extra
dimensions is tested at a much smaller energy scale than the Tev scale
reached in colliders.

Moreover, in high-energy collisions, since the particles are in the
ultra-relativistic regime, the radion exchange has a much weaker contribution
than the graviton exchange in tree-level collisions, since the radion
field couples to the rest mass of the matter \cite{colliders,pankov}.
Therefore the radion is not probed with same precision as the graviton
in those kind of collisions.

As it is known, the radion field is the scalar degree of freedom of
the metric perturbation tensor associated with the trace of the metric
perturbation with respect to the extra directions only ($h_{a}^{a}$).
Its behavior describes fluctuations in the volume of the supplementary
space and it should be subject to some stabilization mechanism for
the extra dimension theory to be viable \cite{stabilization,antoniadis,goldberg,chacko}.
Since this additional mechanism acts only on the radion, it is appropriate
to treat the radion and the graviton as phenomenologically independent
quantities \cite{radion,fabio2023}. In this respect, we can say the
quantum-bouncer experiment has the advantage of being able to probe
the radion with the same precision, since the energy scale $\Lambda$
contains contributions of the same order from the radion and the graviton.

Another important feature to note is that $\Lambda$ depends on the
parameter $\sigma$, and for distinct particles this parameter can
take different values in some versions of thick brane models, which
assume that particles of different nature can be located at different
places in the thick brane \cite{thickbrane}. Therefore, the analysis
of this neutron experiment also contributes to the investigation of
this hypothesis by studying collisions with particles of a different
nature than those usually employed in high-energy colliders.

\section{Final Remarks}

Non-standard scenarios that predict the strengthening of gravity in
short-range scales, such as the large extra-dimensions, are phenomenologically
very interesting and have motivated many table-top experiments to
test the behavior of the gravitational interaction in ever smaller
domains in recent years.

These scenarios are still speculative, but as they are based on solid
theoretical foundations, often related to unification goals, it is
important to search for signals of these exotic theories in various
physics systems, by exploring distinct physical phenomena and by probing
different energy scales as widely as possible.

In this context, experiments involving neutrons are particularly promising,
since the absence of the Coulombian force can be helpful for observing
more clearly possible signs of an anomalous gravitational interaction
between the neutron and matter.

The quantum bouncer is an example of a neutron experiment that has
also been used for this purpose. Deviations from the gravitational
interaction between the neutron and the reflecting mirror have been
analyzed in several papers using the Yukawa \cite{Nesvizhevsky2004,Westphal2}
and power-law parameterization \cite{Fabien}. However, previous studies
on the consequences of a power-law deviation in the gravitational
law have only be able to address scenarios with a small number of
extra dimensions \cite{Fabien}. This limitation results from the
fact that the adopted brane models lead to divergence problems when
calculating the potential energy of the gravitational neutron-mirror
interaction.

In this article, we show that this experiment can be analyzed in more
comprehensive scenarios, where the space can contain an arbitrary
number of additional dimensions, if we take into account the thickness
of the brane. Following this approach, we explicitly found the anomalous
gravitational potential produced by the mirror. From this result,
we were able to identify the physical quantity, relative to the extra
dimensions model, that the present neutron experiment is able to constrain.
This is an effective energy scale, which depends on the gravitational
constant of the ambient space $\left(G_{D}\right)$, and the parameter
$\sigma$, which in turn depends on the profile of the mass-energy
distribution of the nuclei in the transverse direction of the brane.
As we've seen, this parameter in a sense determines a kind of effective
transverse width of matter inside the thick brane.

Although the empirical limits found here are not the strongest compared
to other similar bounds, it is important to note that the empirical
constraints obtained from this neutron experiment are independent
and valid at an energy scale very different from those utilized in
other experiments. Furthermore, our results show that in thickbrane
models the existence of bound vertical motions for ultracold neutrons
in the Earth gravitational field, as verified by this neutron experiment,
is not incompatible with extra-dimensional models with five or more
hidden dimensions.

\section{Acknowledgments}

J. M. Rocha thanks CAPES for financial support.

\end{document}